\documentclass[aps,pra,floatfix,showpacs,twocolumn
]{revtex4}

\usepackage{amsmath}
\usepackage{graphicx}
\usepackage{verbatim}

\newcommand{\bra}[1]{\mbox{$\langle #1|$}}
\newcommand{\ket}[1]{\mbox{$|#1\rangle$}}

\newcommand{\Ket}[2]{|#1\rangle^{(#2)}}

\begin{document}

\title{Loss-tolerant operations in parity-code linear optics quantum computing}

\author{A.~J.~F.~Hayes}
\email{ahayes@physics.uq.edu.au} \affiliation{Centre for Quantum
Computer Technology, University of Queensland, QLD 4072, Brisbane,
Australia.}

\author{A.~Gilchrist}
\affiliation{Centre for Quantum Computer Technology, University of
Queensland, QLD 4072, Brisbane, Australia.}

\author{T.~C.~Ralph}
\affiliation{Centre for Quantum Computer Technology, University of
Queensland, QLD 4072, Brisbane, Australia.}

\date{\today}

\begin{abstract}

A heavy focus for optical quantum computing is the
introduction of error-correction, and the minimisation of resource requirements. We detail a complete encoding and manipulation scheme designed
for linear optics quantum computing, incorporating scalable operations and loss-tolerant
architecture.

\end{abstract}


\maketitle


\section{Introduction}

%
Linear optics is a highly promising architecture in the drive to produce a
quantum computer. It was first shown by Knill, Laflamme and Milburn (KLM) \cite{klm} that linear
optics was a viable system for implementing scalable quantum computing \cite{mikenike}. Further work by various
researchers has produced experimental demonstrations of some of the basic components required by
linear optics quantum computing (LOQC) \cite{PJF03,Obr03,Wal04}. Another focus of work in the field
is on improving the efficiency with which computation could be performed. An alternative scheme put
forward by Nielsen \cite{Nie04} introduced the use of the cluster state model \cite{briegel} in
LOQC. A stream-lined version of this scheme can be found in the paper by Browne and Rudolph
\cite{Bro04}, which significantly decreases the size of the overheads required for computing, when
compared with the original KLM design. More recently, there has been research done on the task of
introducing error-correction into the cluster-state model \cite{0405134, Daw06}. For a more
extensive overview of the field of LOQC, see \cite{Kok07}.

%

We have previously presented an approach to loss-tolerant active memory based on an incremental
parity encoding~\cite{Hay04,Ral05}.  Parity encoding was used in the original KLM proposal to
protect against both teleporter failures (i.e. the non-determinism of the gates) and photon loss.
By using parity encoding but re-encoding incrementally (instead of by concatenation) we are able to
obtain the reduction in overheads characteristic of the cluster state approach whilst retaining the
circuit model and parity encoding of KLM. With the addition of a layer of redundancy encoding, this
allowed for recovery from photon loss.

%
In this paper we will present a universal set of gates for use with a parity-based loss-tolerant
code, to allow scalable quantum computing. We will show that these gates maintain loss-tolerance
during operation, and calculate the loss-tolerant thresholds for computation within the scheme.
Though our techniques for detecting and correcting loss are themselves also subject to loss, above
a particular threshold efficiency the effect of loss can be negated to arbitrary accuracy, making
the computation loss-tolerant.

%
In section \ref{sec:enc} we shall describe the structure of the encoding that allows us to recover
from loss, and the gate operations available to us in designing a system for universal quantum
computation. In this case, we assume the use of photon sources and detectors, linear optical
elements, and fast feed-forward. Section \ref{sec:lto} details the operations that will form a
universal set of gates on the logical qubits. We demonstrate that using re-encoding to perform
these gates allows recovery from losses that occur whilst attempting them. Finally, in section
\ref{sec:ltt}, we calculate the loss threshold for general computation, under this set of
operations. These calculations deal only with loss errors, and do not consider other classes of
error, such as depolarisation. We have focussed on qubit
loss, as it is a dominant source of error in optics, however
it should be noted that by neglecting other
forms of error we are assuming that photon loss is by far
the dominant source of error \cite{Roh07}.

\section{The Encoding}
\label{sec:enc}

%
We will deal with qubits in three different tiers of encoding: (i) \emph{physical} encoding, (ii)
\emph{parity} encoding and (iii) \emph{redundant} encoding. At the first tier are the basic
physical states that we will use to construct qubits, these will be the polarisation states of a
photon so that $\ket{0} \equiv\ket{H}$ and $\ket{1}\equiv\ket{V}$.   The advantage of this choice
in optics, is that we can perform any single physical-qubit unitary \emph{deterministically} with
passive linear optical elements.
Of course gates between different physical qubits become difficult and in LOQC these are typically
non-deterministic. The function of the parity encoding is to allow near deterministic operations
and to convert photon loss to heralded bit-flip errors. The redundant encoding then allows recovery
from these errors.

\subsection{Parity Encoding}

%
We have shown how this class of code may be implemented on an arbitrary number of qubits
\cite{Hay04}. In this paper the notation $\ket{\psi}^{(n)}$ will be used to represent a logical
qubit $\ket{\psi}$  parity-encoded across $n$ distinct physical modes each containing one photon.
We describe these individual photons as the \emph{physical qubits} that make up the system. The
physical qubits also correspond to the first level of encoding.

%
%
A \emph{parity} encoding across $n$ photons is given by
\begin{eqnarray}
\label{parity} \ket{0}^{(n)} & \equiv & (\ket{+}^{\otimes n}+\ket{-}^{\otimes
n})/\sqrt{2}\nonumber \\
\ket{1}^{(n)} & \equiv & (\ket{+}^{\otimes n}-\ket{-}^{\otimes n})/\sqrt{2},
\end{eqnarray}
where  $\ket{\pm} = (\ket{0} \pm \ket{1})/\sqrt{2}$. The $\Ket{0}{n}$ and $\Ket{1}{n}$ states have only even
or odd parity terms respectively. A computational basis measurement of any one
of the physical qubits will merely reduce the level of the parity encoding by one, without losing
the logical qubit. A bit-flip correction may be needed dependent on the measurement result.

%



\subsection{Gates at the Parity Level}

%
%
%
The logical gates described in KLM were based on using concatenation to build up a very large
resource state, and then teleporting the logical qubits in order to apply the gate operation. This
method also allowed for partial loss protection to be built into the gates \cite{thresholds}, but
the resource costs were extremely high. Our alternative scheme \cite{Hay04}, based on the same
code, uses re-encoding to perform gates and has a reduced resource cost as a result.

%
%
The operation that allows us to teleport qubits or entangle states is the partial Bell state
measurement \cite{Wei94,Brau95}. For qubits encoded in the polarisation modes of a photon, this
operation is done by mixing two physical qubits on a polarising beam splitter followed by
measurement in the diagonal-antidiagonal basis. It is successful when one photon is detected in
each arm of the beamspitter's output. If both photons appear at one of the outputs, the operation
has failed. The probability of success for the operation is $1/2$. When successful it projects onto
the Bell states $\ket{00} \pm \ket{11}$, otherwise it projects onto the separable states $\ket{01}$
and $\ket{10}$, measuring the qubits in the computational basis. The operation can be used to
attach physical qubits to a parity encoded state. This is referred to as type-II fusion ($f_{II}$)
\cite{Bro04}.

\begin{figure}
\begin{center}
\includegraphics[width=6cm]{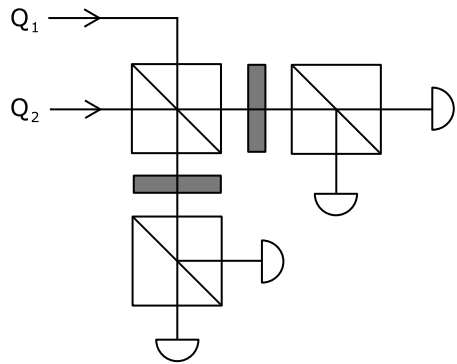}
\caption{This is the optical layout for a type-II fusion gate. It enacts a partial Bell measurement
on two input qubits, acting as an entangling gate with a 50\% probability of success.}
\end{center}
\end{figure}

%
%
There are two operations which are easily performed on parity encoded states. One is a rotation by
an arbitrary amount around the $x$-axis of the Bloch sphere (ie $X_\theta=\cos(\theta/2) I -
i\sin(\theta/2)X$), which can be performed by applying that operation to any of the physical
qubits; and the other is a $Z$ operation, which can be performed by applying $Z$ to \emph{all} the
physical qubits (since the odd-parity states will acquire an overall phase flip). This means that
all the Pauli operations can be performed deterministically. The remaining gates needed in order to
achieve a universal gate set are a $Z_{90}$ and a \textsc{cnot} gate. These can be efficiently
performed on the parity encoded states through re-encoding.


Re-encoding is done by applying a type-II fusion between a physical
qubit from the code state and a resource of $\ket{0}^{(n+2)}$. The
result is
\begin{equation}
    f_{II}\ket{\psi}^{(m)}\ket{0}^{(n+2)}\rightarrow\left\{ \begin{array}{cl}
        \ket{\psi}^{(m+n)} & \mbox{(success)}\\
        \ket{\psi}^{(m-1)}\ket{0}^{(n+1)} & \mbox{(failure)}
    \end{array}\right.
    \label{encoding}
\end{equation}

When this is successful, the length of the parity qubit is extended by $n$ (two qubits are consumed
in the operation). A phase flip correction may be necessary depending on the measurement results.
Failure causes the physical qubit from the parity encoded state to be measured, lowering the level
of encoding by one. The resource state is left in the state $\ket{0}^{(n+1)}$ and can be re-used.

Full details of how to enact the  $Z_{90}$ and \textsc{cnot} gates can be found in Gilchrist {\it
et al.}~\cite{Gil07}.

\subsection{Redundant encoding}

%

%
%

%

%
%
The full loss-tolerant encoding begins with a parity code of length $n$, and concatenates it with a
redundancy code of length $q$. Thus at the highest level our logical qubits are given by:
    \begin{align}
    \ket{\psi}_L &=  \alpha \Ket{0}{n}_1 \Ket{0}{n}_2.....\Ket{0}{n}_q + \beta \Ket{1}{n}_1 \Ket{1}{n}_2.....\Ket{1}{n}_q\nonumber\\
    &= \alpha \bigotimes^{q} \ket{0}^{(n)} + \beta \bigotimes^{q} \ket{1}^{(n)}\nonumber\\
    &= \alpha \Ket{0}{n,q} + \beta \Ket{1}{n,q}\label{lqb}
    \end{align}
where $\bigotimes^{q}$ indicates the tensor product of $q$ such
states.

%
It turns out to be useful to build the following resource state:
\begin{equation}
\ket{0}\ket{0}^{(n,q)} + \ket{1} \ket{1}^{(n,q)}
\label{memRes}
\end{equation}
%
%
We can create an ``encoder'' gate that correctly encodes from a parity qubit to a full redundancy
qubit by simply fusing the resource state above onto the parity qubit. We attempt type-II fusion
between this resource and the parity qubit, $\ket{\psi}^{(n)}$, repeating until successful (on
average twice) giving the (phase flip corrected) result
\begin{align}
& \alpha[\ket{0}^{(n-k)} \Ket{0}{n,q} + \ket{1}^{(n-k)} \Ket{1}{n,q}] + \nonumber\\
& \beta [\ket{1}^{(n-k)} \Ket{0}{n,q} + \ket{0}^{(n-k)} \Ket{1}{n,q}]
\end{align}
where $0<k<n-1$ is the number of unsuccessful attempts made before fusion was achieved. This state
is made up of $q n$ ``new'' photons introduced by the resource and $n-k$ of the ``old'' photons
that made up the parity qubit.  By measuring the old photons in the computational basis and making
a bit flip (on all-new parity qubits if needed) we obtain the expected encoded state (Eq
\ref{lqb}).

\subsection{Active Memory Circuit}

%

The identity operation on the encoded state acts to detect and correct loss errors that may have
occurred. Regularly performing this check can protect the quantum information from loss \cite{Ral05}.

In this operation, one of the constituent parity qubits is sent into the encoder described earlier.
With arbitrarily high probability, the encoder either successfully re-encodes the parity qubit as a
full redundancy state, or it detects a loss. If a loss is detected, measurements in the diagonal
basis $\Ket{0}{n} \pm \Ket{1}{n}$ can be performed on the remaining constituents of the parity
qubit to disentangle it from the rest of the state. Once the logical state is no longer entangled
with the lost photon, the encoding operation may be reattempted.

When the encoder succeeds, diagonal basis measurements can be used to remove the rest of the
original parity qubits from the entanglement. In each case, after disentangling, it may be
necessary to apply a phase-flip to return the logical qubit to the state in Eq \ref{lqb}. Higher
levels of loss can be tolerated by increasing the size of the redundancy code. For a redundancy
code of size $q$, it is possible to tolerate loss on up to $q-1$ of the parity qubits, with the
state being fully re-encoded from the remaining parity qubit.

\section{Logical Gates in Loss Tolerant Encoding}
\label{sec:lto}

%

%
%
To achieve loss-tolerant quantum computing, the next step is to incorporate a full set of universal
gates into the loss-tolerant memory scheme described above. We already have a universal set of
gates at the level of the parity encoding \cite{Gil07}, and these will be the basis for our development of gates
for the loss-tolerant code. The key lies in finding an implementation of a universal set of gates
that can be applied efficiently to qubits in this loss-tolerant encoding.

It is also necessary to ensure that the protection against loss is
not compromised by these operations. As a logical operation
typically consists of a series of gates enacted on physical qubits,
it is possible for losses to occur and be detected during this
process. However, if the component gates have taken the logical
qubit out of the code space, it may no longer be possible to correct
an error. This why it is necessary to design logical operations that
will not compromise the integrity of the code at any point.

%
%
In the parity encoding, we are able to perform arbitrary rotations about the $x$-axis ($X_{\theta}$), $90$-degree
rotations about the $z$-axis ($Z_{90}$), and \textsc{cnot} gates between qubits. These are the fundamental
operations that make up a universal set at that level of encoding. Moving to the redundancy code,
it can be seen that $Z_{\theta}$ rotations on a single parity qubit apply to the entire code, but
that performing $X_{\theta}$ rotations would in general be significantly more difficult. Consequently, we will focus
on implementing ($Z_{\theta}$) and ($X_{90}$) at the redundancy level. However, all the Pauli gates may be
performed deterministically at this level, as at the parity level of encoding.

\subsection{The $Z_{\theta}$ rotation}

%
%
%
Although performing an arbitrary $Z_{\theta}$ rotation on a logical qubit requires merely a
$Z_{\theta}$ rotation on a single parity qubit within the state, such $Z_{\theta}$ rotations on the
parity qubits are not trivial to perform. To enact a $Z_{\theta}$ rotation on a parity qubit using
the set of gates described in \cite{Gil07} would require a couple of steps, during which the logical qubit
is not always in a code state, and hence not properly protected from photon loss. To avoid this
problem, it is necessary to change the procedure for doing an arbitrary $Z_{\theta}$ rotation.

%
%
Consider a general redundancy qubit $\Ket{\psi}{n,q}$:
%
    \begin{align}
    \Ket{\psi}{n,q} &= \alpha \Ket{0}{n,q} + \beta
    \Ket{1}{n,q}\nonumber\\
    &= \alpha \Ket{0}{n,q-1}[\ket{0}^{(n-1)}_A\ket{0}_B+
    \ket{1}^{(n-1)}_A\ket{1}_B]\nonumber\\
    & + \beta \Ket{1}{n,q-1}[\ket{1}^{(n-1)}_A\ket{0}_B+\ket{0}^{(n-1)}_A\ket{1}_B]
    \label{eq:psi}
    \end{align}

We will require a resource state $\ket{R_1}$ to perform a logical $Z_{\theta}$ of the form
    \begin{align}
    &\ket{R_1} = \ket{0}^{n+1} = \ket{0}_C\ket{0}^{(n)} + \ket{1}_C\ket{1}^{(n)}
    \label{parEncRes}
    \end{align}

Step 1 is to perform a $Z_{\theta}$ rotation on a single component qubit of the redundancy state
(qubit $B$):
    \begin{align}
    &\ket{\psi_1} = \alpha \Ket{0}{n,q-1}[\ket{0}^{(n-1)}_A\ket{0}_B+
    e^{i\theta}\ket{1}^{(n-1)}_A\ket{1}_B]\nonumber\\
    & + \beta \Ket{1}{n,q-1}[\ket{1}^{(n-1)}_A\ket{0}_B+
    e^{i\theta}\ket{0}^{(n-1)}_A\ket{1}_B]
    \end{align}

Step 2, a type-II fusion gate is performed between qubit B and a component qubit of the resource
state (qubit $C$). The fusion acts to re-encode the state from the single qubit we have rotated.
    \begin{align}
    &\ket{\psi_2} = \alpha [\Ket{0}{n,q-1}][\ket{0}^{(n-1)}_A\ket{0}^{(n)}+
    e^{i\theta}\ket{1}^{(n-1)}_A\ket{1}^{(n)}]\nonumber\\
    & + \beta [\Ket{1}{n,q-1}][\ket{1}^{(n-1)}_A\ket{0}^{(n)}+
    e^{i\theta}\ket{0}^{(n-1)}_A\ket{1}^{(n)}]\label{psi1}
    \end{align}

Step 3 is to measure in the computational basis the remainder of the parity qubit ($A$). In the event that an odd parity is measured, an $X$ gate on the
newly-added parity qubit is required as a correction.
    \begin{align}
    \bra{0}^{(n-1)}_A&\ket{\psi_2}:\nonumber\\
    \ket{\psi_3} &= \alpha \Ket{0}{n,q-1}\ket{0}^{(n)}
     + \beta \Ket{1}{n,q-1}e^{i\theta}\ket{1}^{(n)}\nonumber\\
    &= \alpha \Ket{0}{n,q} + e^{i\theta} \beta \Ket{1}{n,q}\\
    \bra{1}^{(n-1)}_A&\ket{\psi_2}:\nonumber\\
    \ket{\psi_4} &= e^{i\theta} \alpha \Ket{0}{n,q-1}\ket{1}^{(n)}
     + \beta \Ket{1}{n,q-1}\ket{0}^{(n)}\label{psi2}\\
    X&\ket{\psi_4}:\nonumber\\
    \ket{\psi_5} &=  \alpha \Ket{0}{n,q} + e^{-i\theta} \beta \Ket{1}{n,q}
    \end{align}

It can be seen that the result of an odd parity measurement is a rotation of the form
$Z_{-\theta}$. In this case the logical gate must be re-attempted, using $Z_{2\theta}$. It is worth
noting that the logical $Z_{180}$ operation can be performed deterministically, and hence that a
$Z_{90}$ gate would only need to be attempted once regardless of the outcome of the measurement.

For a general $Z_{\theta}$ gate, an average of two attempts would be required. The advantage this
method holds for our purposes is that the redundancy qubit will always be left in a code state,
maintaining the protection against loss.

\subsection{The $X_{90}$ rotation}
%
%
%
For a universal set of gates, an $X_{90}$ gate is also required. To enact the $X_{90}$ gate, the
operation is performed on one of the component physical qubits. It is then possible to re-encode
from this phyiscal qubit in a similar manner to that used for the $Z_{\theta}$ gate. Measurement of
the old qubits will once again allow us to determine an appropriate set of corrections.

%
%
%
As before, we begin by considering a general redundancy qubit $\Ket{\psi}{n,q}$ (Eq. \ref{eq:psi}).
A larger resource state $\ket{R_2}$ is required for the logical $X_{90}$ gate:
    \begin{align}
    &\ket{R_2} =\ket{0}_C \Ket{0}{n,q} + \ket{1}_C\Ket{1}{n,q}
    \label{redEncRes}
    \end{align}

We then proceed as before, step 1 being an $X_{90}$ rotation on one component qubit ($B$).
Step 2 is to perform a fusion gate between that qubit and the qubit labelled as $C$ in the resource
state.
In step 3, it is necessary to measure all the old qubits which made up the original redundancy
state. Those qubits in the parity state from which the rotated qubit came ($A$) are measured in the
computational basis. All others ($D$) are measured in the diagonal basis. Corrections will depend
on the overall parity of the qubits measured computationally, and on whether an odd number of the
other parity qubits are measured in the $\ket{-}$ state.

The possible states after measurement are:
%
    \begin{align}
     &\bra{0}_A\bra{+}_D\ket{\psi}:\nonumber\\
    &\ket{\psi_6} = (\alpha - i \beta) \Ket{0}{n,q}
     - i (\alpha + i \beta) \Ket{1}{n,q}\nonumber\\
     &\bra{1}_A\bra{+}_D\ket{\psi}:\nonumber\\
    &\ket{\psi_7} = (\beta - i \alpha) \Ket{0}{n,q}
     - i (\beta + i \alpha) \Ket{1}{n,q}\nonumber\\
     &\bra{0}_A\bra{-}_D\ket{\psi}:\nonumber\\
    &\ket{\psi_8} = (\alpha - i \beta) \Ket{0}{n,q}
     + i (\alpha + i \beta) \Ket{1}{n,q}\\
     &\bra{1}_A\bra{-}_D\ket{\psi}:\nonumber\\
    &\ket{\psi_9} = (\beta - i \alpha) \Ket{0}{n,q}
     + i (\beta + i \alpha) \Ket{1}{n,q}\nonumber
    \end{align}

Accordingly, we may require a logical $X$ gate, a logical $Z$ gate, or both in order to correct the
resulting state. Once any necessary corrections are performed, we are left with a redundancy state
on which the $X_{90}$ operation has been successfully applied.

\subsection{Logical \textsc{CNOT}}

It was explained earlier in this paper that a logical \textsc{cnot} gate could be enacted on a
parity qubit by a process of encoding. The \textsc{cnot} is performed between two redundancy
qubits, $\ket{\psi}$ and $\ket{\phi}$. Here $\ket{\psi}$ is the control and $\ket{\phi}$ is the
target.
    \begin{eqnarray}
    \ket{\psi}_L = \alpha \Ket{0}{n,q} + \beta \Ket{1}{n,q}\\
    \ket{\phi}_L = \gamma \Ket{0}{n,q} + \delta
    \Ket{1}{n,q}
    \end{eqnarray}
The logical \textsc{cnot} gate is performed as an iterative process, with a parity-level
\textsc{cnot} performed for each parity qubit in $\ket{\psi}$. Each of these parity-level
\textsc{cnot} gates will use an arbitrary parity qubit from $\ket{\phi}$ as its target input.

We use the following resource for each iteration.
    \begin{eqnarray}
    \ket{R_3} = \ket{0}_C \ket{0}^{(n)}(\ket{0}^{(m)}\ket{0}_D+ \ket{1}^{(m)}\ket{1}_D)\nonumber\\
    + \ket{1}_C \ket{1}^{(n)}(\ket{1}^{(m)}\ket{0}_D+ \ket{0}^{(m)}\ket{1}_D)
    \end{eqnarray}
where $m=\lfloor n/2\rfloor$.  It consists of two parity qubits with a \textsc{cnot} already performed between them. The target parity qubit, $m$, is shorter since re-encoding is not required on the second logical qubit.

Step 1 of the process is then to fuse a member of the selected parity qubit from $\ket{\psi}$ with
qubit $C$ in the resource, $\ket{R_3}$.

If this is successful, step 2 is to measure the remaining original physical qubits in the parity
state, to complete the re-encoding. If a loss is detected anywhere up to this point, we disentangle
the chosen parity qubit and the resource from the rest of the $\ket{\psi}$ state using diagonal
basis measurements. This allows us to recover and re-attempt the process.

In step 3, perform a fusion gate between qubit $D$ in the resource and a physical qubit taken from
$\ket{\phi}$. To recover from a loss, should one occur during this fusion, we disentangle the
resource from $\ket{\psi}$ by making diagonal basis measurements on it, and do the same for the
parity qubit in $\ket{\phi}$ on which we have acted. Once again, $Z$ and/or $X$ gates may be
required as corrections on both qubits depending on the outcome of the measurements. These Pauli
gates can be applied deterministically to parity or redundancy states.
To perform the full logical \textsc{cnot}, this process is iterated for each parity qubit in
$\ket{\psi}$.

%
%
These encoding-based gate operations continually replace the photons used for the logical qubits,
and the old photons are measured, identifying any losses that arise. In this way, loss detection
and correction are continually applied during computation.

\section{Loss-Tolerant Thresholds}
\label{sec:ltt}

In order for the encoding to be useful in a scalable quantum computing scheme, it is necessary to
show that a loss threshold exists. If the loss is below this threshold, it is possible to
drive the probability of failure arbitrarily close to zero by increasing the size of the code. We
will first summarize the threshold calculation for the identity operation, as presented in our
previous paper \cite{Ral05}. We then present a revised threshold for general computation, using the
logical gate operations we have described.

\subsection{A Loss-Tolerance Threshold for the Active Memory}

%
%
The active memory scheme is used to protect an encoded logical qubit $\Ket{\Psi}{n,q}$, by regularly re-encoding it using the resource given in Eq~\ref{memRes}.
We begin by considering the probability of loss for each photon. The efficiency of the photon
source will be labelled $\eta_s$, and the efficiency of the detectors will be $\eta_d$. We will use
$\eta_m$ to indicate the memory efficiency, which is the probability a photon will \emph{not} be
lost during the time it is in memory, in-between re-encoding cycles. This means that the
probability of detecting a new photon, from a resource state, is $\eta_2=\eta_s\eta_d$, and the
probability of detecting an old photon, from the code state, is $\eta_1=\eta_s\eta_m\eta_d$. Note
that fusing a resource onto a logical qubit will succeed or fail with probability $\eta_1\eta_2/2$
and detect a photon loss with probability $1-\eta_1\eta_2$.

In calculating the loss-tolerant threshold for the encoding, we consider the possible outcomes of
an attempt to re-encode the state. For a given parity qubit in the overall state, there are three possible outcomes when attempting to re-enode from it.

The first possible outcome is successful re-encoding without loss. This occurs when the fusion is successful on one of the first $n-1$ physical qubits in the parity state, and the remainder are measured in the computational basis without loss. The probability for this is:
\begin{equation}
P_{Qs} = \sum_{i=1}^{n-1}(\frac{1}{2} \eta_1 \eta_2)^i \eta_1^{n-i}
\end{equation}
Note that if only one component qubit remains in the parity state, we instead measure it in the diagonal basis to disentangle it, and begin again with another parity qubit from the overall state.

The second outcome that can occur is total failure. This can result from a long series of losses and/or fusion failures. The probability for total failure is:
\begin{multline}
P_{ff} = \sum_{j=1}^{n-1}(\frac{1}{2} \eta_1\eta_2)^{j-1} (1-\eta_1\eta_2)(1-\eta_1)^{n-j} \\
+ R\sum_{j=0}^{n-2}(\frac{1}{2} \eta_1\eta_2)^{j+1}\sum_{k=0}^{n-2-j}\eta_1^{k} (1-\eta_1)^{n-1-j-k} \\
+(\frac{1}{2} \eta_1 \eta_2)^{n-1}(1-\eta_1)
\end{multline}
\begin{equation}
R=\sum_{k=1}^{q}{q \choose k}(1-\eta_2)^{kn}[1-(1-\eta_2)^{n}]^{q-k}
\end{equation}
Here $R$ is the probability of failing to recover via measurements on the new resource qubits. This can occur when photon loss is detected after a fusion has been performed, and attempts to disentangle by measuring components of the original parity qubit have proven unsuccessful.

The third possibility is that of recovery after partial failure. The probability of this can be calculated from the previous two equations: $P_{Qf} =
1-P_{Qs}-P_{ff}$. We can tolerate this outcome occurring up to $q-1$ times when attempting to re-encode. Therefore, the total probability for successfully re-encoding is:
\begin{equation}
P_{E} = \sum_{j=0}^{q-1} P_{Qf}^{j} P_{Qs}[1-(1-\eta_1)^n]^{q-1-j}
\end{equation}
where the $[1-(1-\eta_1)^n]^{q-1-j}$ factor occurs because it is necessary to disentangle the remainder of the original state once the chosen parity qubit has been successfully re-encoded.

\begin{figure}
\begin{center}
\includegraphics[width=8cm]{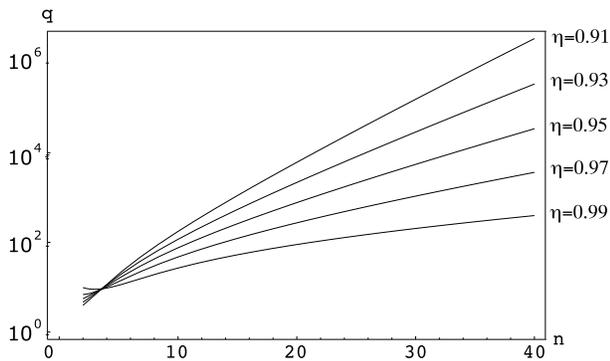}
\caption{Variation of the optimal value of $q$ with $n$, for given values of $\eta$}
\label{ratio}
\end{center}
\end{figure}

\begin{figure}[h]
\begin{center}
\includegraphics[width=8cm]{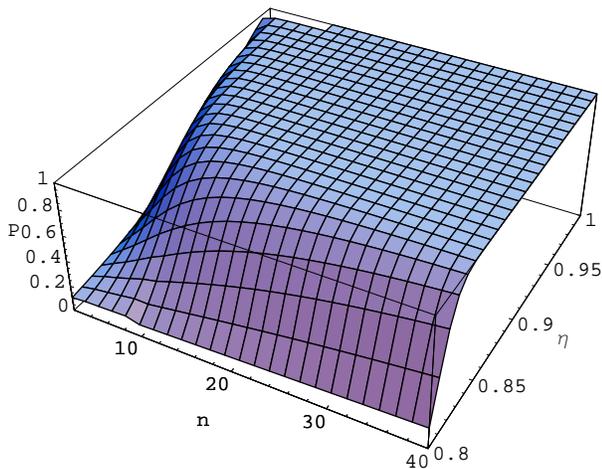}
\caption{Probability of success for active memory using optimal $q$ as a function of $\eta $ and
$n$}
\label{limits}
\end{center}
\end{figure}

For the probability $P_E$ to approach one for large encodings, it is necessary to maintain a particular ratio between $n$ and $q$. The optimal ratio can be found by solving $\frac{d}{dq}P_{E}=0$ for $q$ in terms of $n$ This relationship is shown in figure~\ref{ratio}.
In these calculations, we considered an equally high error rate in all parts of the circuit ($\eta_s = \eta_m = \eta_d = \eta $). Using the optimal ratio, we found numerically that $P_E$ can be driven arbitrarily close to one when $\eta \geq 0.82$. This is shown in figure~\ref{limits}.


\subsection{A Loss-Tolerance Threshold for Computation}

%
%



The thresholds for the single-qubit logical gates are the same as the threshold for the identity
(memory) case, due to the strong similarity between the gate operations, and the re-encoding used
in the active memory. The $Z_{\theta}$ operation uses smaller resources, and as such has a slightly
higher probability of success, but this difference becomes vanishingly small for large code sizes.
This results in it having the same threshold as the $X_{90}$ operation, and the identity.
The \textsc{cnot} gate is the most complicated of the universal set of gates we have developed, and
we would expect it to be the most vulnerable to loss.



The probability of successfully performing our \textsc{cnot} gate without loss can be evaluated by
considering three possible outcomes for each iteration in the procedure. These consist of a
no-progress outcome, in which a \textsc{cnot} between parity qubits fails due to loss or
measurement errors, a progress outcome, in which the \textsc{cnot} is successful, and total
failure, in which one or both logical qubits are lost.
There are several ways in which a no-progress outcome can occur. These events and their probability
are listed below.

1. Fusion attempts are unsuccessful, and the chosen parity qubit is disentangled from the rest of
the state:
\begin{equation}
M_1=(\frac{1}{2} \eta_1 \eta_2)^{n-1} \eta_1
\end{equation}

2. A loss occurs during a fusion attempt, and the parity qubit is disentangled:
\begin{equation}
M_2=\sum_{i=0}^{n-2}(\frac{1}{2} \eta_1 \eta_2)^{i} (1-\eta_1 \eta_2) [1-(1-\eta_1)^{n-i-1}]
\end{equation}

3. A loss occurs while measuring off qubits after a successful fusion, and the parity qubit is
disentangled:
\begin{align}
&M_3=\\&\sum_{i=1}^{n-2}(\frac{1}{2} \eta_1 \eta_2)^{i} \sum_{j=0}^{n-i-2} \eta_1^j
(1-\eta_1)
 [1-(1-\eta_1)^{n-i-j-1}]\nonumber
\end{align}

4. A loss occurs while measuring off qubits after a successful fusion, the parity qubit is not
disentangled, and it is necessary to measure the resource in order to disentangle it:
\begin{align}
M_4=\sum_{i=1}^{n-1}(\frac{1}{2} \eta_1 \eta_2)^{i}& \sum_{j=0}^{n-i-1} \eta_1^j
(1-\eta_1)^{n-i-j}\\&[(1-(1-\eta_2)^{n}) (1-(1-\eta_2)^{\frac{n}{2}+1})]\nonumber
\end{align}

5. A loss occurs during fusion with the target qubit, which is measured in order to disentangle it:
\begin{equation}
M_5=\sum_{i=1}^{n-1} (\frac{1}{2} \eta_1 \eta_2)^{i} \eta_1^{n-i} (1-\eta_1) (\frac{1}{2} \eta_1
\eta_2) (1-(1-\eta_1)^{\frac{n}{2}+1})
\end{equation}

For most of these events, it is necessary to re-encode the logical control qubit afterwards to
ensure it is fully protected. This has a probability of success of $P_E$.

Hence the probability of a no-progress outcome ($M$) is:
%
%
\begin{equation}
    M=P_E \sum_{k=1}^4 M_k + M_5
\end{equation}
%
%
%
%
Here $P_E$ is the probability of successfully re-encoding a logical qubit, as shown earlier. The
probability of a progress outcome ($K$) is:
\begin{multline}
    K= \sum_{i=0}^{n-1}(\frac{1}{2} \eta_1 \eta_2)^{i} \eta_1^{n-i} [\frac{1}{2} \eta_1 \eta_2 \\+ (1-\eta_1 \eta_2) (1-(1-\eta_1)^{n-1})
    (1-(1-\eta_2)^{\frac{n}{2}})]
\end{multline}

For a gate between two logical qubits, each made up of $q$ parity qubits, the overall probability
of success is:
\begin{equation}
    P_{TOTAL}= \left(\frac{K}{1-M}\right)^q
\end{equation}

\begin{figure}[h]
\begin{center}
\includegraphics[width=8cm]{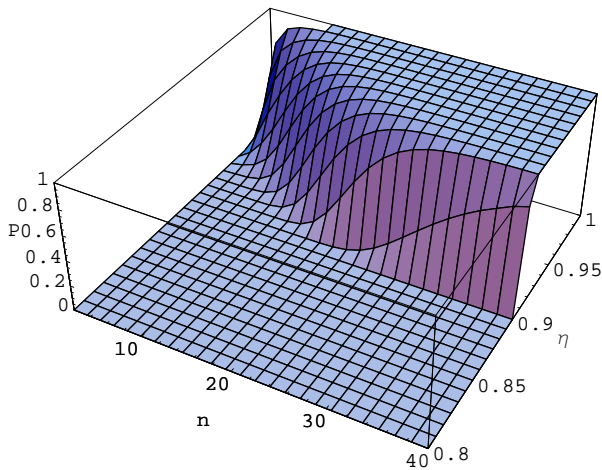}
\caption{Probability of success for a logical \textsc{cnot} gate on redundancy qubits}
\label{cngatep}
\end{center}
\end{figure}

%
%
To simplify the calculation, we again consider the case in which the different parts of the system
(sources, memory/manipulation, detectors) contribute equally to the loss. We represent the
efficiency of each of these components as $\eta$. Using the equations shown, we can examine the way
the probability of success varies with this efficiency (figure~\ref{cngatep}). It is assumed the
$n:q$ ratio of these logical qubits will be optimised for re-encoding, using the formula found
earlier.

It can be seen that the threshold approaches a value of 90\% efficiency for the \textsc{cnot} gate.
So far, we have assumed an equal contribution to the loss from different components. However, if we
assume that one or more parts of the system are lossless (e.g. perfect detectors), the thresholds
for the rest of the system will drop accordingly.


%
%

\section{Resources}
\label{sec:res}

The procedures we have described require many entangled resource states to be prepared separately
for use in computation. We assume our basic building blocks for these resources to be maximally-entangled
Bell pairs, in the state $\Ket{0}{2}$. Such Bell pairs would have to be generated directly from a
heralded source, or created from single photons via a KLM-style entangling gate \cite{klm}.

The first step in creating the resources required is to generate larger parity states, of the form
$\Ket{0}{n}$. These states are built up iteratively by fusing smaller parity states together, in a
similar manner to that used to generate cluster states \cite{Bro04}. Initially, it is necessary to
use a type-I fusion gate, which acts as a single-rail partial Bell measurement. As in the case of
the type-II fusion gate, both input qubits are mixed at a 50-50 beamsplitter. However, only one arm
is measured. The type-I gate is successful when exactly one photon is found in this arm. To achieve
the desired fusion operation for combining parity states, Hadamard gates are performed on the
inputs and output of the type-I fusion gate. This operation has the advantage that only one qubit
is measured, but a failure means that both parity states are completely lost.
\begin{align}
    (H\!\otimes\! H) f_I H \Ket{0}{n}\Ket{0}{m}&\rightarrow \left\{ \begin{array}{cl}
        \Ket{0}{m+n-1} & \mbox{(success)}\\
        - & \mbox{(failure)}
    \end{array}\right.\label{eq:fIjoin}
\end{align}

As a result, the type-I fusion gate is used to create short parity qubits, which are then joined in
larger chains using type-II fusion \ref{encoding}. The type-II fusion measures one qubit from each
state, but does not destroy the state in the event of failure, allowing resources to be recycled.
Additionally, if one of the input qubits to the type-II gate is missing the loss will be detected,
reducing loss errors due to gates in the entanglement construction.
\begin{figure}[h]
\begin{center}
\includegraphics[width=8cm]{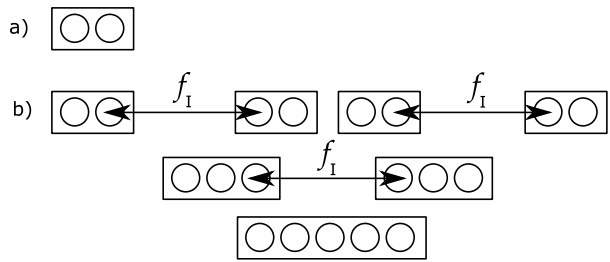}
\caption{(a) A representation of the two-qubit parity state used as a starting resource,
$\Ket{0}{2}$. (b) The process by which larger parity states are generated, using fusion gates
between qubits.} \label{figres1}
\end{center}
\end{figure}
For efficient resource production, states of the form $\Ket{0}{5}$ could be produced using type-I
fusion, as shown in figure \ref{figres1}. This would require an average of 16 Bell pairs. In order
to generate larger states, the type-II fusion gate should be used to join multiple copies of the
$\Ket{0}{5}$ state.

To create redundantly encoded resources, we begin by performing a \textsc{cnot} between a pair of
physical qubits taken from the states $\Ket{0}{n+1}$ and $\Ket{0}{n}$. If successful, this will produce the state
\begin{equation}
\ket{0}\Ket{0}{n,2}+\ket{1}\Ket{1}{n,2} \label{eq:q2res}
\end{equation}
which can be used to encode a qubit \ket{\Psi} in the logical state $\Ket{\Psi}{n,2}$. To build a
larger redundancy resource, we fuse multiple copies of the state given by Eq. \ref{eq:q2res}.
\begin{figure}
\begin{center}
\includegraphics[width=5.8cm, height=5.4cm]{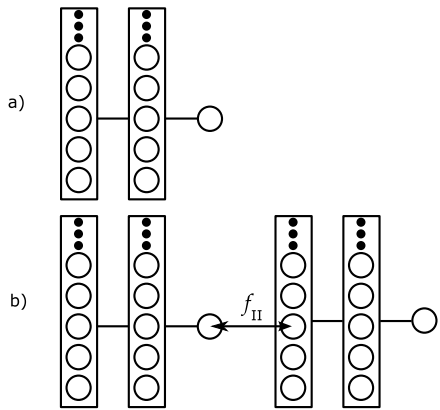}
\caption{(a) A graphical representation of the state $\ket{0}\Ket{0}{n,2}+\ket{1}\Ket{1}{n,2}$
(equation \ref{eq:q2res}). (b) A fusion between two such states. Once the remainder of the parity
qubit being encoded has been measured in the computational basis, the resulting state is
$\ket{0}\Ket{0}{n,3}+\ket{1}\Ket{1}{n,3}$.}
\end{center}
\end{figure}

\section{Conclusions}


In this paper we have given a full description of a redundancy code for performing circuit-based linear optical quantum
computing in the presence of photon loss. We have found that the
code is successful as a quantum memory if each potential area of loss (the photon source, the
memory/operational section of the circuit, and the detectors) has an efficiency of 82\% or greater.
For general computation in this system, the threshold efficiency is 90\%, as this is the minimum
efficiency which will allow all the gate operations to work successfully. For both of these
thresholds, an efficiency less than the threshold in an area of loss can be tolerated if other
areas have correspondingly higher efficiencies. We have restricted our consideration to 
photon loss which has enabled us to describe a quite complete error 
correction protocol for general quantum computation with a high loss 
tolerant threshold. However, neglecting other noise sources is rather 
unrealistic. In future work we plan to address more general error 
correction codes based on optical parity states.

\section*{Acknowledgments}
This work was supported by the 
Australian Research Council, Queensland State Government, and 
DTO-funded U.S. Army Research Office Contract No.
W911NF-05-0397.

\bibliography{references}

\end{document}